\documentclass[applsci,article,submit,pdftex,moreauthors]{mdpi} 

\usepackage{amssymb}

\firstpage{1} 
\pubvolume{1}
\issuenum{1}
\articlenumber{0}
\pubyear{2023}
\copyrightyear{2023}
\datereceived{ } 
\daterevised{ } 
\dateaccepted{ } 
\datepublished{ } 
\hreflink{https://doi.org/} 

\Title{Trust model of privacy-concerned, emotionally-aware agents in a cooperative logistics problem}

\TitleCitation{Trust model of privacy-concerned, emotionally-aware agents in a cooperative logistics problem}

\Author{Javier Carbo $^{1,\dagger,\ddagger}$\orcidA{} and Jose Manuel Molina $^{2,\ddagger}$ \orcidB{}}

\AuthorNames{Javier Carbo, and Jose Manuel Molina}

\AuthorCitation{Carbo, J.; Molina, J.M.}

\address {%
$^{1}$ \quad Computer Science Dept., Univ. Carlos III de Madrid; javier.carbo@uc3m.es\\
$^{2}$ \quad Computer Science Dept., Univ. Carlos III de Madrid; molina@ia.uc3m.es}

\corres{Correspondence: jcarbo@inf.uc3m.es}

\firstnote{Current address: Av Gregorio Peces Barba 22, Campus de Colmenarejo, Madrid.} 
\secondnote{These authors contributed equally to this work.}

\abstract{
In this paper we propose a trust model to be used into a hypothetical mixed environment where humans and unmanned vehicles cooperate. We address the inclusion of emotions inside a trust model in a coherent way to the practical approaches to the current psychology theories. The most innovative contribution is how privacy issues play a role in the cooperation decisions of the emotional trust model. Both, emotions and trust have been cognitively modeled and managed with the Beliefs, Desires and Intentions (BDI) paradigm into autonomous agents implemented in GAML (the programming language of GAMA agent platform) that communicates using the IEEE FIPA standard. The trusting behaviour of these emotional agents is tested in a cooperative logistics problem where: agents have to move objects to destinations and some of the objects and places have privacy issues. The execution of simulations of this logistic problem shows how emotions and trust contribute to improve the performance of agents in terms of both, time savings and privacy protection}

\keyword{Trust; Emotion; Agents; Privacy}

\begin{document}

\section{Introduction}  \label{sec:1}

Including emotions in the interactions between humans and autonomous computational entities (agents) is one the current challenges of Artificial Intelligence:  the so called affective computing \cite{picard1997}. Success in this challenge would provide efficiency and common understanding in such interactions. Trust (how trust is achieved, applied and updated) is also of great relevance in how and whether such interactions between humans and agents take place. Both very human concepts, trust and emotion, have been largely addressed in the field of psychology from different theoretical approaches in the scientific literature as we will see in section \ref{sec:1b}. We intend to suggest a system of agents that integrates the existing link between Privacy, Trust and Emotions in a meaningful way that is coherent with such literature. 

We intend to achieve this goal inspiring ourselves with a hypothetical cooperative logistics problem where humans and embodied unmanned vehicles gait according to their internal emotions, and where they also perceive the emotions of the others they meet (by observing their gait). Although our simulation does not include humans neither such futuristic embodied unmanned vehicles which are able to emotionally gait, the implemented agents reason and communicate through a human-like way, and they perceive a simulated emotion from an hypothetical gait of the agents they meet. In such suggested cooperative logistic problem, these agents have to perform repetitive tasks (moving boxes). Since several of these tasks may overwhelm the ability of an agent, they request the cooperation of other agents to perform some of these tasks (move boxes) in behalf of them. Asking and accepting the cooperation involve forming a trusting relationship, where trust in other agents is built not only with the knowledge of the past direct interactions with them, trust is also decided with the internal emotions of the agent and with the interpretation of the perceived emotion of the interacting agent. In this way, internal emotions and perception of alien emotions of the other agent act as indirect knowledge in the trusting decision replacing the role that reputation information about third parties often play in trust models. The trust model we suggest is therefore focused on social punishment applied to agents misbehaving (non cooperative behavior) through the reasoning about own and alien emotions (acquired by observation and by the privacy concern associated to the task to accomplish). 

On one hand, our intended hypothetical case of use consists of humans and embodied unmanned vehicles which are both moving boxes in the same environment. But nowadays the existing unmanned vehicles either are able to autonomously transport boxes in the real urban environment or they are embodied and able to gait as humans, but both features are currently exclusive. In the case of only humans having the ability to gait then emotions are to be perceived just from them and our model have to be significantly adapted to this limitation. 

On the other hand, the privacy issues of the boxes may take several forms depending on the nature of the box itself, for instance some examples of real life situations where humans may feel shame are those regarding to the potential object they are carrying:
\begin{itemize}
\item A bouquet of flowers, specially if the handler is seen in a suspicious way due to personal circumstances: already married, too young/old, etc.
\item A stroller for a baby, specially if their work colleagues do not know about it.
\item A set of masks when there is shortage of them in a pandemic situation.
\item A piece of cloth not coherent with the perceived gender role.
\item Any kind of item with a political/ideological meaning.
\end{itemize}
The possible reasons that justify privacy issues inducing shame in the handler are of very different nature and it would be nearly impossible to produce an exhaustive list of them, our work is not intended to be specific for any of them. 

Our contribution is not exclusively focused on the tasks to be accomplished in time (in our case of use of the logistic problem, sharing the boxes to be moved would produce time savings), it is also strongly linked to the avoidance of private disclosures. The avoidance of these privacy disclosures can significantly contribute to decrease the negative effects of social prejudices on discriminated minorities. The combined goal we pursue consists of both of them, and the experiments result we will show in section \ref{sec:6} are driven to show how the emotions and the trust model influence both.

Specifically, the implementation of the agents of our contribution takes the form of an humanized internal architecture that includes emotions and privacy reasoning, which is an internal cognitive/symbolic representation of human reasoning where such agents interact between them through an implicit observation of the gait of the others they meet and the explicit exchange of human-like messages asking for cooperation. 

Section \ref{sec:1b} illustrates the state of the art corresponding to the computational representation of Emotions, Trust and Privacy. Section \ref{sec:2} describes how we address the definition and computation of emotions and personality in our approach. Section \ref{sec:3} shows the trust model applied and its relationship with emotions and personality. Section \ref{sec:4} describes the FIPA protocols used and the BDI reasoning followed by agents. Section \ref{sec:5} shows the cooperative logistic problem defined to test the execution of our model using the GAMA agent platform. Section \ref{sec:6} includes the experimental results. And finally, section \ref{sec:7} concludes discussing the possible uses and benefits of our model.  

\section{State of the Art} \label{sec:1b}

\subsection{Computational representation of Emotions}

There is no consensus in the academic literature on how emotions can be categorized or how emotions arise and apply \cite{johnson2009}. The simplest classification distinguishes six basic/primary/innate emotions \cite{ekman1999} (anger, disgust, fear, joy, sadness and surprise) or eight ones \cite{plutchik2003} (trust, fear, surprise, sadness, disgust, anger, anticipation and joy). However when emotions are limited to facial expressions, \cite{beckeretal2007} reduced them to four instances:  anger, sadness, fear, joy and surprise. A more fine grained representation \cite{damasio1994} included secondary emotions which result from the evaluation of the expectations such as relief or hope. The adoption of these emotions into an affective state (called mood) involve at least one dimension, the pleasure it generates (called valence) that can be either positive (good/pleasant) or negative value (bad/unpleasant). The most complete categorization is the (so called OCC by the name of its authors) cognitive theory \cite{occ}, which distinguish emotions according to the source: events, agents and objects. A human/agent is on a particular mood when it feels the emotion with enough intensity, which would decrease over time. Therefore, many formalizations \cite{russell80} \cite{watson85} include at least a second dimension, noted as arousal, that represents the (positive) level of intensity/excitement regarding to the given emotion. But the most widely used model \cite{mehrabian1974} includes a third one: dominance and it is known as PAD (Pleasure, Arousal and Dominance) model. It represents how much control feels the human/agent facing the current situation. A positive value would mean the human/agent is dominant (in control of the emotion), and a negative one would mean the human/agent is submissive. In the PAD model, an emotion is then represented by a point in a three dimensional space. A mood is adopted when a given emotion exceeds a given threshold, and it decreases over time (a forgetting curve) depend on the personality of the human/agent. Unlike mood, personality is not temporal and does not depend on the occurrence of a particular event/context. The most extended theory \cite{mccrack92} about personalities considers five factors: openness, conscientiousness, extroversion, neuroticism and agreeableness, although a so popular personality theory \cite{eysenk67} just considers three factors: extroversion, neuroticism and psychoticism. Extroversion is linked to more display, sensitivity and duration of positive emotions while neuroticism has similar effect but to negative emotions \cite{hwee}. Additionally, neuroticism shows a reduced ability to focus attention to complete tasks \cite{Rothbart2011} jointly with aversion to novelty and uncertainty \cite{KaganFox06}. Psychoticism traits show rejection of cultural norms and non social compliance to the social expectations \cite{Eysenck76}. 

Emotions arise in an uncontrolled way (the submissive ones) producing sudden embodied effects that may be observed physically \cite{ledoux1996}  \cite{damasio1994}. There are several previous anthropomorphic systems that employ fully embodied agents to show emotions, for instance in physical interactions \cite{casselletal2000} \cite{prendinger2004}, in conversations \cite{rosisetal2003} \cite{ochsetal2006}, and overall the most frequently used method of showing emotions is through facial expressions \cite{breazeal2003} \cite{itohetal2006}. Although less used as a way to show and perceive emotions, the gait of humans and embodied agents has also been studied \cite{Roether} \cite{Xu2022}. Additionally, automated detection of human emotions has raised considerable privacy and legal concerns \cite{disclosure} \cite{emotional} that are not addressed in this contribution.

\subsection{Computational representation of Trust}

Agents are intended by its own nature to be self-interested, their behaviour can not be assumed to be altruistic and cooperative \cite{weiss}\cite{WooldridgeyJennings1995}. Since they are also dynamic, past behaviour of an agent can only be a rough estimation of its future behaviour. Additionally agents systems are also by definition open, so agents come and go, and therefore different levels of knowledge about the behaviour of other agents coexist, and finally as they can be also highly populated, the frequency of own possible interactions with a majority of the other agents in the system is not likely at all. The joint existence of these features produces a significant difficulty in estimating the expected reliability of potential partners in interactions. 
In order to overcome such difficulty, trust models based on reputation computations are being proposed along the last decades \cite{sabater2005}. While reputation stands for a quantitative evaluation of the expected behaviour of another agent which has to be numerically aggregated and updated from indirect sources and direct experiences (each model suggests a different computation), trust stands for the cognitive decision of an agent to participant in a potentially risky or uncertain interaction \cite{castellfranchi}. 

Cognitive decisions of agents are often implemented using the Beliefs, Desires and Intentions (in advance, BDI) paradigm, which produce agent intelligent behaviour through cyclic deliberation about explicit beliefs, desires and intentions of the agent \cite{rao}. Such cognitive agents are expected to make use of interaction protocols to communicate meaningful messages with each other following the IEEE  Foundation for Intelligent Physical Agents (in advance, FIPA) standard \cite{fipa}. Such trusting cognitive decision in BDI agents considers factors such as the beliefs about the general situation in which the interaction decision is going to take place and the beliefs about the other agents \cite{barberchallenges}. While in real life, emotions play a key role in trusting decisions, the beliefs agents use do not represent emotions, they decide with the beliefs of the agent about the situation and about the other agents unemotionally. Even the only properly emotional trust model we found in the literature \cite{emotionaltrust} decides with a trust model and an emotional model implemented as independent reasoning blocks. However, the need of an emotional trust model for agents has been recently recognized in the literature \cite{carles}. Furthermore, \cite{ocsrevisited} observes a close relationship between the beliefs of agents and the emotions described by the OCC model of emotions \cite{occ}. And finally, it should be noted that underlying some of the purely trust models proposed, such as \cite{singhyu} where failures in the trusting decision have much more mathematical impact in reputation than success, there is a hidden implicit emotional process.               

\subsection{Computational representation of Privacy issues in social interactions}

Privacy protection has been largely bounded by legal systems, specially in the European Union (EU), as Regulation 2016/679  of the European Parliament shows \cite{directive2016}. While Privacy can be roughly defined as the right of controlling the own personal information \cite{thenet}, it is is a very complex and broad issue that exceeds the legal perspective. Even if legal (since they may take place in public spaces), some interactions cause the perception of potentially shameful, intimate or sensible tasks by the interacting partners. The nature, scope, context and purpose of any social interaction defines the level of privacy harm produced \cite{wright2013making} \cite{stewart2012privacy}. The acknowledge of this level of potential privacy harm in the own and alien social image of the interactions causes an emotional response in both interacting parts \cite{stark}. Therefore, as in the real life, privacy not only plays a legal role, it also plays an emotional role in the interactions (inhibiting or promoting behaviour). In the same way, when agents act autonomously on behalf of humans, the emotional impact of the potential privacy-related interactions performed by the agent have to be represented, weighted and considered in the decision making of such agent. The particular cause of the privacy issue involved in the social interactions may vary, they can be related to ideological, gender, work and health issues. And they can take very different forms according to the personal circumstances that take places in the social interactions. It is not the intention of this contribution to be focused in any one of them, or to specifically represent a model of all these causes and their given circumstances that launch the privacy concern. We just assume that social meetings sometimes have associated privacy issues that could be avoided through cooperation between humans and autonomous agents. In our case of use of the logistic problem when a human delegates the move task of a box to an unmanned vehicle, and therefore avoiding privacy issues of the box itself or the places to move through later whatever these privacy issues are, and whatever the causes that produce them.   

\section{Proposed Emotional model} \label{sec:2}
\subsection{Sources of emotions} \label{sources}
Among all the possible representations of emotions described in section \ref{sec:1}, we decided to represent a set of 5 emotions out of the 6 basic ones of \cite{ekman1999}: joy, sadness, anger, fear and surprise. This decision is caused by the limitation of how emotions are assumed to be perceived and showed in our contribution (through the gait of humans and embodied agents). The different gait perceptions of agents that agents might for instance perceive when they meet are:
\begin{itemize}
\item happy: for instance, when the met agent is walking straight and looks front-side.
\item sad: for instance, when the met agent is curving the back and looking to the floor.
\item anger: for instance, when the met agent is rising its shoulders, its hands are forming a fist, and its moves are fast and rigid.
\item fear: for instance when the met agent becomes standstill, and its hands and legs start shaking.
\item surprise: for instance when the met agent becomes standstill, and its head goes back while it rises its hands. 
\end{itemize} 

In our model, in order to represent the emotive reactions that take place in real life, agents feel emotions from the combination of three different dynamic external perceptions (the sources of emotion):
\begin{enumerate}
\item the others: When the perceived emotion to the agent currently meeting is anger, a fear emotion will be produced.
\item alien privacy: When there is a privacy issue of the other agent involved in the meeting (which can be due to the the object the other agent is carrying), a surprise emotion will be produced. 
\item own privacy: When there is a privacy issue of our agent involved in the meeting (which can be due to the the place of the meeting or the object our agent is carrying), an anger emotion will be produced.
\item the positive rewards of our performance: When our agent is successfully accomplishing its task, joy will be produced
\item the negative rewards of our performance: When our agent is badly accomplishing its task, sadness will be produced.
\end{enumerate}
Additionally, we suggest agents to have one of the three personalities proposed by \cite{eysenk67}: extroversion, neuroticism and psycoticism. The personality an agent has acts as emotion-enhancer: While extroversion personality tends to enhance positive emotions (joy, surprise), neuroticism enhances negative emotions (sadness, fear) and psycoticism enhances antisocial emotions (anger). While emotions are dynamic and dependent of the sequential situations the agents address and several emotions may coexist (although just one of them is dominant and becomes the mood of the agent), personality is static, predefined and mutually exclusive.

\subsection{The PAD levels of the emotions}

\begin{figure}
\centering
\includegraphics[width=12cm, height=6cm]{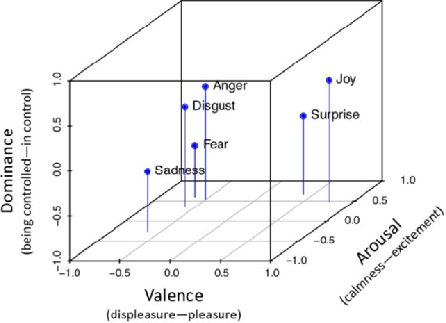}
\caption{3D representation of the basic emotions as shown in \cite{Buechel2016}}
\label{fig:emo3d}
\end{figure}

From the combination of these sources, in each iteration cycle, a resulting value will represent the possibility of an agent to feel an emotion. Since several emotions may simultaneously being felt, the emotion corresponding to the greater of them is the one that the agent shows when two agents meet in the same place. As the literature (section \ref{sec:1}) states, emotions are often defined by three dimensions: pleasure, arousal and dominance (PAD), the 5 basic emotions related to gait perception have associated the specific values shown in table \ref{tbl:pad} \cite{Russell1977}. Due to the intrinsic meaning of these concepts, while pleasure and dominance take positive and negative values (between -1 and 1), arousal is always positive (between 0 and 1). We can graphically observe from figure \ref{fig:emo3d} how the basic emotions are placed in a 3 dimensional space as \cite{Buechel2016} showed.
The feeling of an emotion ($w_{e}$)is computed using the equation \ref{eq:feelemo} where $d_{e}$ stands for the distance between the current PAD 3d position of the agent and of each basic emotion, $\Delta_{e}$ is the minimum threshold to activate an emotion and $\phi_{e}$ establishes the point of saturation of each emotion as \cite{becker-asano2010} proposed.

\begin{table}[h!]
\caption{PAD definition of each emotion according to \cite{Russell1977}}
\centering
\begin{tabular}{ | c | c c c | }
\hline
Emotion & Pleasure & Arousal & Dominance \\ 
\hline
 joy  & 0,75 & 0,48 & 0,35 \\
 sad & -0,63 & 0,27 & -0,33 \\
 surprise & 0,4 & 0,67 & -0,13 \\
 fearful & -0.64 & 0,6 & -0,43 \\
 angry & -0.51 & 0,59 & 0,25 \\
\hline 
\end{tabular}
\label{tbl:pad}
\end{table}

\begin{equation}
\label{eq:feelemo}
    w_{e}= \left( 1 - \frac{d_{e} - \Delta_{e}}{\phi_{e} - \Delta_{e}} \right) 
\end{equation}
 
Since emotions tend to be balanced along time, we progressively reduce the pleasure level of an emotion $P_{i}$ with a decreasing function that depends on how excited is the agent (the arouse level) in the way showed by equation \ref{eq:balancep} where $V_p$ is a constant that softens the level of decreasing fixed in $0.1$.

\begin{equation}
\label{eq:balancep}
    P_{i}=P_{i-1} - V_p \times A_{i-1}
\end{equation}

\subsection{PAD changes due to the source of emotions}

Each of the 5 sources of emotion of subsection \ref{sources} (others, own privacy, alien privacy, positive and negative rewards) are intended to cause the below explained emotional reaction (fear, anger, surprise, joy and sadness), but through changes in the PAD values.
 
Pleasure level is positively affected (with a value of 0.1) by the instant satisfaction that an agent feels when a reward for achieving a current goal is satisfied (in our logistic problem, when the box reaches the destination in time). On the other hand, when the agent fails achieving a current goal (in our logistic problem, it takes place in both situations: social punishment due to own privacy disclosure and failure in the box delivery), its pleasure level becomes negatively affected in the opposite way (with a value of -0.1).  

In order to compute the level of arousal of the agent, we propose that it increases in a directly proportional way to the sum of the variation of the next potential source of emotions: the perceived anger emotion of the other agent currently meet and the both privacy disclosures (own and alien) that are induced by the boxes and the place of the meeting. Equation \ref{eq:arousal} shows this computation, where $V_{a}$ is a constant that softens the level of increasing/decreasing fixed in $0.1$ and where $S_{i,j}$ is the contribution to the arousal for each j sources of emotions. 

\begin{equation}
\label{eq:arousal}
    A_{i}=A_{i-1} + \sum{ {S_{i,j}} - {S_{i-1,j}}} \times V_a
\end{equation}

Due to the emotions we intend to produce from the sources of emotion, dominance becomes decreased (by 0.1) when anger is perceived in the gait of the agent is currently meeting. But also since dominance represents the dichotomy between being controlled and in control, two other conditions cause a sense of control/discontrol in the agent:
\begin{itemize}
\item The number of tasks which the agent is in that moment responsible of (in our logistic problem, due to own and delegated boxes to move): Since a high number of them causes a sense of discontrol of the situation is currently addressing. Each task exceeding the first one causes a decrease of -0.05 in the dominance, on the other hand, having no current task causes an increase of 0.1 in the dominance, and just one task a 0.05 increase. 
\item The overall achievement/performance of the agent along all the execution (not just an instant reward for the current goal): Since a bad performance causes sense of discontrol (it enforces the agent to accept the requests of help from untrusted agents), the accumulated reward per task value less than the average one causes a decrease of -0.1 in the dominance, while a greater value would cause an increase of 0.1.   
\end{itemize}

As we did with pleasure, a decreasing function is applied over the dominance component. The level of decreasing also depends on how excited is the agent (the arouse level itself) in the way showed by equation \ref{eq:balanced} where $V_d$ is a constant that softens the level of decreasing fixed in $0.1$.

\begin{equation}
\label{eq:balanced}
    D_{i}=D_{i-1} - V_d \times A_{i-1}
\end{equation}

\subsection{Influence of personality into emotions and performing ability}

As the literature states that personality influences how emotions are felt in different ways: extroversion enhances joy and surprise, neuroticism enhances sadness and fear and psychoticism enhances anger. We intend to represent such influence through the way changes in pleasure, arousal and dominance are computed. Specifically, we suggest to use the constants, $V_p$, $V_a$ and $V_d$ that softens the level of decreasing over time for pleasure, arousal and dominance. The resulting modifications are:
\begin{itemize}
\item Since joy and and surprise have associated the bigger positive values of pleasure in table \ref{tbl:pad}, extroversion personality will promote joy and surprise (as \cite{hwee} suggested) by causing less level of changes in pleasure (through a smaller softening constant ($V_p = 0.05$ instead of $V_p = 0.1$) when it is positive.  
\item Since sadness and and fear have associated the bigger negative values of pleasure in table \ref{tbl:pad}, neuroticism personality will promote sadness and fear (as \cite{hwee} suggested) by causing less level of changes in pleasure (through a smaller softening constant ($V_p = 0.05$ instead of $V_p = 0.1$) when it is negative.  
\item Since anger has associated a big positive values of dominance in table \ref{tbl:pad}, psychoticism personality will promote anger by causing less level of changes in dominance (through a smaller softening constant ($V_d = 0.05$ instead of $V_d = 0.1$) when it is positive.   
\end{itemize}

The accomplishment of the moving tasks is strongly associated to the personality of the agents. Since neuroticism causes a reduced ability to focus attention to complete tasks according to \cite{Rothbart2011}, all (own and delegated) moving tasks are delayed (by a cycle) when performed by neurotic agents. Since psychoticism leads to rejection of cultural norms and non compliance to the social expectations according to \cite{Eysenck76}, psychotic agents perform the delegated moving tasks with a delay of a cycle. 
 
\section{Proposed Trust model} \label{sec:3}

The interactions between agents cause both rewards and punishments that strongly influence the way agents trust each other. On one hand, the privacy concern caused by other agents (in the logistic problem, whenever our agent meets other agent in a sensible cell or with a sensible box) produces a negative feedback (social punishment) to our agent. On the other hand, our agent will receive positive feedback (social reward) when the task is successfully accomplished (in the logistic problem, when the box reaches the destination). If moving such box was partly delegated to other (trusted) agents, our agent would receive a reward corresponding to the level of joint success of the trusted agents proportionally to the incurring delays in the delivery. So we can now distinguish two different rewards that agents receive: the privacy-related and the delay-related. While both of them have two subtypes: related to the box and place the privacy one, and related to the own and alien performance the delay-related. 

We can distinguish two different trusting decisions: 
\begin{enumerate}
\item The decision to request cooperation, where our agent becomes the trusting agent since it delegates a task into other agent (the trusted one) taking some risks (in our logistic problem, trusting the other agent to move a box towards its destination). There is no associated certainty or guarantee in the future behaviour of the other agent, although afterward the trusting agent will receive a delayed feedback about the behaviour of the trusted agent (in our logistic problem, the trusting agent will know whether the box reached the destination in a given time or not). 
\item The decision to answer a cooperation request from other agent, where our agent becomes the trusting agent since it carries out a delegated task for other agent (the trusted one) taking some risks (in our case, trusting the other agent to reach the destination in a given time). Again, there is no associated certainty or guarantee in the future behaviour of the other agent, although afterward the trusting agent will receive a delayed feedback about the behaviour of the trusted agent (in our logistic problem, the trusting agent will know whether the moving task was performed in time or not).
\end{enumerate}

The decision to request cooperation from other agent will depend upon the next criteria:
\begin{itemize}
\item The mood (current feeling) of the trusting agent. Positive emotions (joy, surprise) of the agent will encourage the trusting decision, while negative (sadness, fear) and antisocial (anger) emotions will discourage the trusting decision with bonus/malus of 0.1 of the trust required.  
\item The privacy issues involved for the agent in the trusting decision (in our logistic problem, the level of privacy associated to the box to be delegated). If privacy issues are involved for us, then we require 0.1 less of trust to the other agent. 
\item How much the other agent is trusted, where trust is computed from the previous performance of the other agent with us (in our logistic problem, the level of success of the other  agent previously performing our moving tasks in the past): A success (delivery of the box without delay) means an increase of 0.1 in trust, while each cycle of delay means -0.05 of decrease in trust.
\item How much the agent needs help (in our logistic problem, the number of already carrying boxes). For each already carrying box, we require 0.05 less of trust to the other agent.   
\end{itemize}

The decision to answer a cooperation request from other agent will depend upon the next criteria:
\begin{itemize}
\item The mood (current feeling) of the trusting agent. Positive emotions (joy, surprise) of the agent will encourage the trusting decision, while negative (sadness, fear) and antisocial (anger) emotions will discourage the trusting decision with bonus/malus of 0.1 of the trust required.
\item The privacy issues involved for the agent in the trusting decision (in our logistic problem, the level of privacy associated to the box to be moved). If privacy issues are involved for us, then we require 0.1 less of trust to the other agent. 
\item How much the other agent is trusted, where trust is computed from the previous performance of the other agent with us (in our logistic problem, the level of success of the other agent previously moving our boxes in the past): A success (delivery of the box without delay) means an increase of 0.1 in trust, while each cycle of delay means -0.05 of decrease in trust.
\item How much the other agent may help (in our logistic problem, the number of already carried and assigned boxes with their time requests and relative paths to their destination). For each already carrying box, we require 0.05 more of trust to accept the cooperation offer.
\end{itemize}

We can remark that the perceived emotion of the other agent does not directly influence both trusting decisions, the influence is indirect: perceived alien emotion influences the own emotion, and such own emotion influence the corresponding trusting decision. In the same way, privacy issues of the to-be-trusted agent does not directly influence both trusting decisions, they influence the alien emotion, which influences the own emotion, and such own emotion influence the corresponding trusting decision.

\section{FIPA protocols and BDI reasoning} \label{sec:4}

According to the BDI paradigm \cite{Rao95}, the behaviour of agents is determined by the achievement of desires, through the execution of the plans corresponding to the intended desire that are fired by the perception of certain conditions (beliefs expressed in the way of predicates). We will see for the agents in our system the adopted desires, the firing beliefs and the corresponding plans they use.

The agents of our system perform their moving tasks following a repeated iteration cycle. Before such cycle takes place, agents are initiated with:
\begin{itemize} 
\item A random fixed personality chosen between the three possible ones: neurotic, psychotic and extroverted.
\item An initial mood (current feeling) derived from the neutral PAD values pleasure 0, arousal 0 and dominance 0.
\item An initial location (random chosen anywhere in the existing grid)
\item An initial desire of being $idle$.
\end{itemize}

Agents are continuously perceiving the existence of the boxes which are assigned to them, jointly with the destination they have to reach and whether they have privacy issues or not. In the same way, agents are continuously perceiving other agents which are busy, jointly with their location. These perceptions cause the adoption of the beliefs $pending package$ and $busy carrier$ respectively, and they occur in parallel to any desires the agent have. 

Once initialized, the repeating iteration cycle starts with the execution of the desire to be $idle$. The corresponding plan is accomplished following the next steps if any pending package is perceived:
\begin{itemize} 
\item Transforming the closest package of all the pending ones into a $current package$ belief.
\item Transforming all other pending packages into $moving package$ beliefs.
\item Dropping the $idle$ desire and adopting the $moving$ desire.
\end{itemize}
In case of no pending packages were perceived, the agent moves towards the location of the closest $busy carrier$ (where busy means currently moving at least one package).

The corresponding plan of the desire of $moving$ a package also transforms all pending packages into $moving package$ beliefs. Afterward, in case of another agent present in the same cell, and any moving package additional to the current package carries out the next steps:  
\begin{itemize} 
\item Chooses the farthest package as candidate to be delegated
\item Recovers the trust in the agent met in the same cell
\item Computes the trust modifiers from the emotions and personality
\item Decide to ask for delegating the candidate package in case of enough trust in the agent met.
\end{itemize}

\begin{figure}
\centering
\includegraphics[width=10cm, height=8cm]{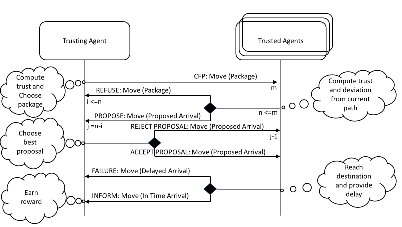} 
\caption{FIPA CFP Interaction Protocol for delegating a task}
\label{fig:cfp}
\end{figure}

The decided delegating process is implemented as a call for proposal (in advance, CFP) interaction protocol (FIPA-compliant), and it is graphically outlined in figure \ref{fig:cfp}.

\begin{figure}
\centering
\includegraphics[width=10cm, height=8cm]{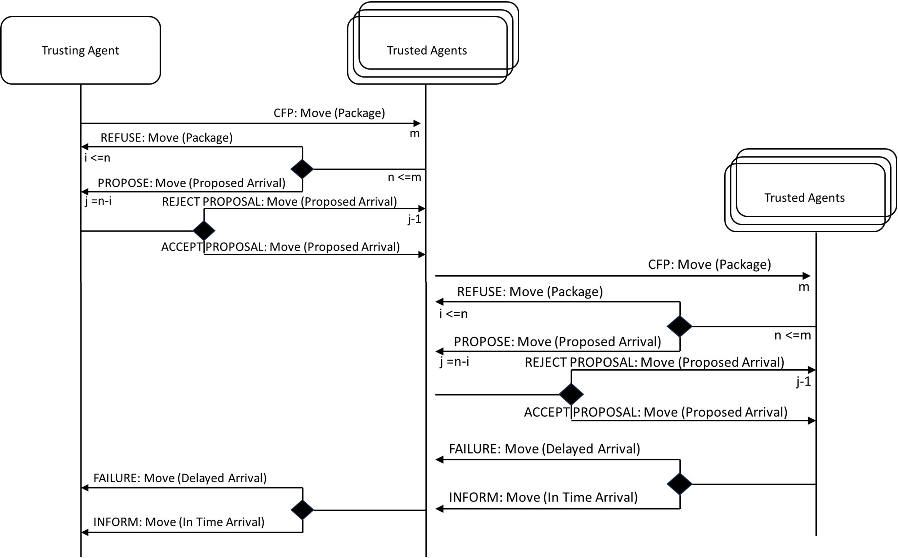} 
\caption{FIPA CFP Interaction Protocol sequentially linked for delegating several times for several agents a given task}
\label{fig:cfpsequence}
\end{figure}

This delegation can takes place several times for the same package, forming sequential instances of this cfp protocol. Figure \ref{fig:cfpsequence} shows graphically such linked sequence of cfp protocol instances.

Once a package has been moved to its destination, it disappears from the simulation.

\section{Problem definition in GAMA platform} \label{sec:5}

In the simulations, carrier agents (with different initial random location, personality and assigned packages) are going to move and interact in an abstract environment represented by a grid. In the beginning of the simulation, several boxes appear with random destinations and assignment to carriers. Initially and in any time, an agent may have concurrently several boxes to move (requiring the cooperation of other agents to satisfy all the moving goals in time). The simulation ends when all the packages have reached their destination. 

Through this problem definition, agents will feel different emotions (when they meet each other and when they succeed or fail moving the box). We assume agents perceive seeing gait of the other agents when they meet in the same location. Additionally we assume the ability of agent to identify which boxes and cells cause privacy concerts to itself. 

In order to implement our trust model we used the GAMA agent platform \cite{gama}. It is an open source software which is FIPA compliant and includes the possibility of using the BDI paradigm in the  reasoning \cite{gamabdi}. GAMA allows large scale simulations and integrates  geographic information systems (in advance, GIS) data \cite{gamagis}. There is also an extension that links norms with emotions for this platform \cite{gamaemotions}. 

\begin{figure}
\centering
\includegraphics[width=6cm, height=8cm]{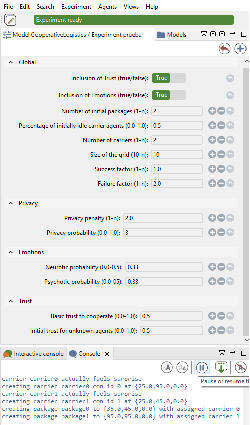}
\caption{Parameter definition in the simulation GUI}
\label{fig:parameters}
\end{figure}

The parameters of the simulation define the setup of the problem as it is shown in figure \ref{fig:parameters}, they are listed below jointly with the values used in the experimentation:
\begin{itemize}
\item Whether to include trust in the simulation or not (variable: true and false)
\item Whether to include emotions in the simulation or not (variable: true and false)
\item Number of packages (fixed: 15)
\item Number of carrier agents (fixed: 15)
\item Percentage of initially idle carrier agents (variable: 0\%, 20\%, 40\%, 60\% and 80\%)
\item Size of the square grid, in number of cells (fixed: 30)
\item Probability of a cell/box to be private (variable: 0.0, 0.2, 0.4, 0.6 and 0.8)
\item Probability of being neurotic (fixed: 33\%)
\item Probability of being psychotic (fixed: 33\%)
\item Penalty associated to privacy disclosure (fixed: 2.0)
\item Reward for reaching the target in time (fixed: 1.0)
\item Penalty for a delay reaching the target (fixed: 2.0)
\item Basic trust threshold to cooperate (fixed: 0.5)
\item Initial trust for an unknown agent (fixed: 0.5)
\end{itemize}

\begin{figure}
\centering
\includegraphics[width=10cm, height=6cm]{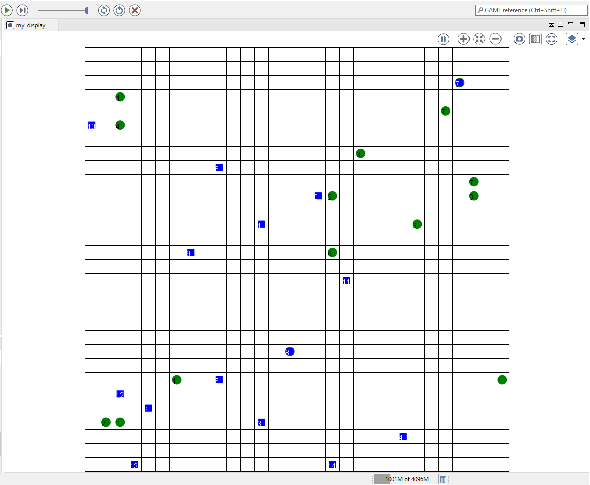}
\caption{Initial situation in a scenario where carrier agents will move to the destination of their boxes}
\label{fig:initial}
\end{figure}	

\begin{figure}
\centering
\includegraphics[width=10cm, height=6cm]{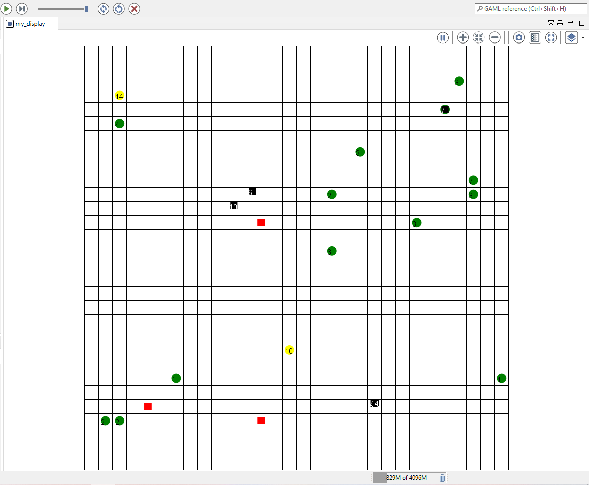}
\caption{Ongoing situation in a scenario where carrier agents are moving to the destination of their boxes}
\label{fig:moving2}
\end{figure}

Agents represented as rectangles with their id inside in a grid of cells, colored in blue when they are idle while the destination of the boxes to be moved by the original assignment are represented with a green-colored circle (with the id of the assigned carrier inside) as it is shown in figure \ref{fig:initial}. Once the agents perceive the packages and starts moving, the destination of the boxes to be moved by other agent different from the original assignment would be represented with a yellow-colored circle (with the id of the currently moving carrier inside), the carriers moving a package are represented by a black-colored square (with its id inside) and  cells where two or more agents meet are represented as red-colored squares as it is shown in figure \ref{fig:moving2}. 

\section {Experimentation results} \label{sec:6}

As any experimentation is going to be strongly dependent on the particular values that the many variables of this model has, the scope of this experimentation is far away from proving any definitive validation. Our goal is to show a possible way of how privacy-induced emotions may integrate in a trust model making possible a futuristic interaction of unmanned automated elements with humans both transporting objects. The role of emotions is not specifically designed to improve the results of the rewards obtained by the agents and although it affects these rewards, they just intend to mimic human behaviour and react as realistically as possible as a human does following the psychology theories and the previous practical approaches to these theories that were explained in section \ref{sec:1}.
  
Using the problem definition of section \ref{sec:5}, we will compare two alternatives with our emotional trust model (noted as $emotionaltrust$): 
\begin{itemize}
\item A trust model without emotions (noted as $noemotions$), this means that the current feeling would not modify how much trust is required to propose and to accept a proposal of a delegation of a task (being happy and surprised decreases it, while being sad, anger or fearful increases it).  
\item No trust model at all (noted as $notrust$), so no agent trust any other, and no cooperation takes place (no objects are delegated to other agent to be carried towards the destination at all). All agents carry the initially assigned objects to their destination by themselves, without any way to decrease the corresponding delays. This is alternative shows the worst case, the benchmark to show the room of improvement achieved by the other alternatives. 
\end{itemize}

\begin{figure}
\centering
\includegraphics[width=10cm, height=6cm]{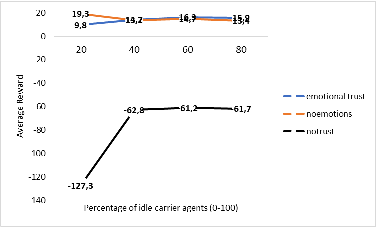}
\caption{Comparison of the 3 alternatives when the percentage of initially idle carrier agents increases}
\label{fig:idle}
\end{figure}

All the comparisons were repeated 100 ways to decrease the variability caused by the initial random locations of objects and agents. The comparison will take place measured in two ways. First, we will see how these three models obtain rewards when the number of idle agents changes. When the number of initially idle agents increases, the possibility of cooperation increases (less delays and more privacy issues avoidance becomes possible too). Figure \ref{fig:idle} shows the results obtained in this first comparison. From the observation of this figure we can conclude that:
\begin{itemize}
\item We could expect the rewards for any alternative to increase, as the percentage of idle agents increases, but it appears to reach a saturation point between 20\% and 40\% and no significant increase is afterward perceived.
\item With independence of the percentage of idle agents, the notrust alternative obtains much less rewards than the other two alternatives (noemotions and emotionaltrust).
\item When the percentage of idle agents is very low (20\%), there are just a few boxes to delegate and causes significant different results to those obtained with greater percentages. This is specially true for the notrust alternative.
\item Except when the percentage of idle agents is very low (20\%), the use of emotions slightly increases the rewards obtained by agents (emotionaltrust line is slightly over noemotions line). So it appears that in these cases, using emotions in the trust model causes some improvement. But the difference is very little.
\item When the the percentage of idle agents is very low (20\%), the use of emotions clearly obtains worse rewards, it seems to decrease the quality of the decisions taken by the trust model (emotionaltrust line is definitively below the noemotions line)
\end{itemize}
 
\begin{figure}
\centering
\includegraphics[width=10cm, height=6cm]{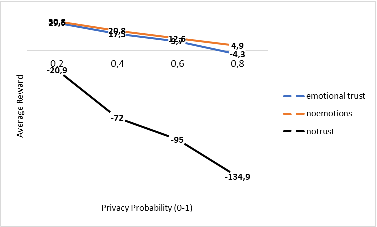}
\caption{Comparison of the 3 alternatives when the probability of privacy issues increases}
\label{fig:privacy}
\end{figure}

In the second comparison we will see how these three models obtain rewards when the privacy probability changes. Privacy probability stands for the probability of a box or a cell to become a privacy issue for a particular agent. The privacy of the box is constant and fixed from the moment the assignment took place (both initially and for delegation), while the privacy of a cell is computed whenever a meeting takes place. For both of them, the same probability value is used. Therefore, in one extreme, a zero privacy probability would mean no cell nor box would be private anytime. So privacy would play no role in the simulation. As we are going to study the effect of the progressive increase in this variable, we remark here the role of this variable in the model:
\begin{itemize}
\item Currently carrying a private box would increase the chances of proposing a delegation to other agent whenever a meeting takes place since the box may be not private for the other agent (and therefore less burden for the other agent to carry it to its destination).
\item Both, private boxes and cells, cause a punishment in the reward whenever a meeting takes place.
\item Both, private boxes and cells, cause a decrease in the pleasure level of the agent (PAD variable) whenever a meeting takes place.
\item Both, private boxes and cells, cause a increase in the arousal level of the agent (PAD variable) whenever a meeting takes place.
\end{itemize}
Figure \ref{fig:privacy} shows the results obtained in this second comparison. From the observation of this figure we can conclude that:
\begin{itemize} 
\item As it was expected, as privacy probability increases, the rewards of any alternative decrease.
\item With independence of the value of privacy probability, the notrust alternative obtains much less rewards than the other two alternatives (noemotions and emotionaltrust).
\item Except when the privacy probability is very high (0.8), the use of emotions slightly decreases the rewards obtained by agents (emotionaltrust line is slightly below noemotions line). So it appears that in these cases, using emotions in the trust model causes some decline. But the difference is very little.
\item When the the privacy probability is very high (0.8), the use of emotions appears to obtain worse rewards, it seems to decrease the quality of the decisions taken by the trust model (emotionaltrust line is below the noemotions line)
\end{itemize}

Taken the results of both comparisons jointly, we observe that the inclusion of emotions in a trust model as we did, following the psychology theories and the previous practical approaches to these theories that were explained in section \ref{sec:1} and with the particular values of all the variables of our model, does not universally improve the rewards obtained just by the trust model without emotions. In some circumstances, it slightly improves the quality of the taken decisions, in some others it clearly declines such quality. In most of them, the differences are very little, and comparing to the no use of a trust model, emotions do not cancel the most significant advantages (in rewards) that a trust model provides.  

\section {Conclusions} \label{sec:7}

Interactions between autonomous agents and humans in mixed environments face several challenges, although most of the focus is set on conversational means, emotions (their expression, perception and reasoning) may also play a major role. We have addressed the topic of using emotions inside a trust model with an innovative inclusion: the role of privacy issues in an emotional trust model. To the best of our knowledge, this has not been addressed before in the scientific literature. To reach such goal, our research has accomplished several milestones: 
\begin{itemize} 
\item We have proposed a particular way to include privacy in an emotional model which is compliant to the psychology theories and the previous practical approaches to these theories that were explained in section \ref{sec:1}. 
\item We have proposed a particular way to include privacy in the cooperation decisions of a trust model.
\item We have suggested a set of particular values to the all set of variables that form our privacy-sensible emotional model. 
\item We have implemented in GAML (the programming language of GAMA agent platform) this model into symbolic agents reasoning according to the Believes, Desires and Intentions deliberative paradigm, that communicates using the IEEE FIPA standard.   
\item We also defined a cooperative logistic problem to test our model.
\item And finally we have executed agent simulations that generated two different comparisons. Such comparisons allowed us to observe the contribution of emotions and trust in the defined cooperative logistic problem to improve both our goals: time savings and privacy protection. 
\end{itemize}

In our proposal, agents intend to be as humanized as possible through a symbolic approach that provides explainability and transparency through the use of a BDI reasoning with FIPA communications. Privacy issues are also an important factor in how we feel interacting with others, and emotional agents should address them in their perceptions, expression and reasoning. The inclusion of privacy sensitivity in an emotional trust model is so innovative than comparison no comparison with alternatives obtained by different researchers is even possible.

Although the experimental results we obtained from the simulation of a logistic problem do not seem to encourage the use of emotions to satisfy our both goals (time savings and privacy protection), the inclusion of emotions is assumed to be a key element for autonomous agents to gain acceptance in real life applications. And since the emotions do not harm significantly the results, its inclusion could balance the little loss of performance with the improvement of representativeness that emotions provide to social interactions. This is the most significant contribution of our work rather than the potential gain that sharing tasks itself can provide. 

Our contribution is not the only way to model the social punishments privacy issues cause, and many particular settings could take another form and lead to very different experimental results, from the conceptual representation of emotions going by the problem used to test the model and finally to the mathematical quantification of all the involved factors. Many other alternative uses of privacy in an emotional interaction between humans and autonomous agents are possible, but our model provides a step in a way worth to be explored.

The adaptation of our emotional trust model to the intended hypothetical mixed environment of humans and unmanned vehicles is out of the scope of this contribution since it would require taking into account the existence of embodied unmanned vehicles expressing emotions through their gait that we now only can imagine. The uncertainty and the range of the perception of the different type of autonomous vehicles would have to be considered. Also other circumstances would have to be taken into account as the capacity and range of the different type of autonomous vehicles. Other transportation issues as the traffic, weather conditions or the different relevance of the boxes may also play a role. The real involvement of humans in the experiment would also make even more difficult to test the model and reach any conclusion. But the usefulness of this research is not dependent of such adaptation, our model is useful because it provides a first proposal on how to involve privacy in the emotional deliberation that takes place in trust decisions. In spite of the lack of specificity in the definition of what boxes are carrying the agents, of what causes privacy concerns, and of the multiple adaptations required to apply it in real life, our proposal show how privacy can be related to the emotional representation of agents trusting each other, and how, when emotions and trust are applied, the agents moving boxes of our simulation address the combined goal of time saving and privacy protection.

\vspace{6pt} 

\funding{This work was funded by public research projects of Spanish Ministry of Science and Innovation (CACTUS), reference PID2020-118249RB-C22 and Spanish Ministry of Economy and Competitivity (MINECO), reference TEC2017-88048-C2-2-R. This work has also been supported by the Madrid Government under the Multiannual Agreement with UC3M in the line of Excellence of University Professors (EPUC3MXX), and in the context of the V PRICIT (Regional Programme of Research and Technological Innovation).}

\dataavailability{Data sharing not applicable to this article as no datasets were generated or analyzed during the current study. 
The complete code of all the agents jointly with the environment and setup files are publicly available in the Sourceforge repository \footnote{https://trustemotionalagents.sourceforge.io} in order to provide transparency and to facilitate the complete replicability of all the simulations included in this manuscript.}





\reftitle{References}

\bibliography{bibs}




\PublishersNote{}
\end{document}